\documentclass[%
 reprint,
superscriptaddress,
amsmath,amssymb,
aps,
pra,
]{revtex4-1}
\usepackage{caption}
\usepackage{subcaption}
\usepackage{graphicx}
\usepackage{xcolor}
\usepackage{graphicx}
\usepackage{dcolumn}
\usepackage{bm}

\begin{document}

\preprint{APS/123-QED}

\title{Refraction of space-time wave packets: II. Experiments at normal incidence}

\author{Alyssa M. Allende Motz}
\affiliation{CREOL, The College of Optics \& Photonics, University of Central Florida, Orlando, FL 32816, USA}
\author{Murat Yessenov}
\affiliation{CREOL, The College of Optics \& Photonics, University of Central Florida, Orlando, FL 32816, USA}
\author{Basanta Bhaduri}
\affiliation{CREOL, The College of Optics \& Photonics, University of Central Florida, Orlando, FL 32816, USA}
\author{Ayman F. Abouraddy}
\thanks{corresponding author: raddy@creol.ucf.edu}
\affiliation{CREOL, The College of Optics \& Photonics, University of Central Florida, Orlando, FL 32816, USA}




\begin{abstract}
The refraction of space-time (ST) wave packets offers many fascinating surprises with respect to conventional pulsed beams. In paper (I) of this sequence, we described theoretically the refraction of all families of ST wave packets at normal and oblique incidence at a planar interface between two non-dispersive, homogeneous, isotropic dielectrics. Here, in paper (II) of this sequence, we present experimental verification of the novel refractive phenomena predicted for `baseband' ST wave packets upon normal incidence on a planar interface. Specifically, we observe group-velocity invariance, normal and anomalous refraction, and group-velocity inversion leading to group-delay cancellation. These phenomena are verified in a set of optical materials with refractive indices ranging from 1.38 to 1.76, including MgF$_2$, fused silica, BK7 glass, and sapphire. We also provide a geometrical representation of the physics associated with anomalous refraction in terms of the dynamics of the spectral support domain for ST wave packets on the surface of the light-cone.
\end{abstract}

\maketitle

\section{Introduction}

Space-time (ST) wave packets \cite{Yessenov19OPN} are a unique family of pulsed optical beams whose spatio-temporal spectra are endowed with a particular structure that dictates their group velocity independently of the refractive index of the medium \cite{Reivelt03arxiv,Kiselev07OS,Turunen10PO,FigueroaBook14}. Specifically, the spectral support domain of an ST wave packet on the surface of the light-cone lies at its intersection with a tilted plane, such that each spatial frequency is associated with a single wavelength \cite{Donnelly93PRSLA,Kondakci16OE,Parker16OE,Kondakci17NP,Yessenov19PRA}. The `spectral tilt angle' of this plane dictates the wave-packet group velocity \cite{Yessenov19PRA,Yessenov19OE}. Because refraction in general produces changes in the spatio-temporal structure of the optical field, one anticipates novel refractive behavior to emerge for such unique field configurations \cite{Bhaduri20NP}. In paper (I) of this sequence \cite{Yessenov21JOSAA}, we studied theoretically the refraction of ST wave packets at a planar interface between two non-dispersive, homogeneous, isotropic dielectrics, and a host of fascinating refractive phenomena were brought to light, even at normal incidence \cite{Bhaduri20NP}.

Refraction relates to the change in the propagation direction of an optical beam across an interface between two media. When a pulsed field is incident normally at such an interface, the transmitted pulse travels in the same direction beyond the interface at a group velocity determined by the local optical properties of the medium \cite{SalehBook07}. Because ST wave packets have a finite spatial bandwidth, a finite range of incident angles must always be taken into consideration, even when the wave packet is nominally incident perpendicularly onto the interface. Although this of course applies equally to any optical beam of finite transverse spatial extent, the spatio-temporal structure characteristic of ST wave packets produces unique refractive consequences that are absent from conventional wave packets in which the spatial and temporal degrees of freedom (DoFs) are separable \cite{Yessenov19OPN}.

In general, the transverse momentum (spatial frequency) and the temporal frequency are invariant across a planar interface. Because each spatial frequency in a ST wave packet is associated with a particular temporal frequency, these conservation rules produce changes in the structure of the spatio-temporal spectrum of the transmitted wave packet, which then induces a change in its group velocity \cite{Bhaduri20NP,Yessenov21JOSAA}. Uniquely, this group velocity no longer depends solely on the refractive index of the second medium, but depends also on the group velocity of the incident wave packet and the refractive index of the first medium. That is, the transmitted wave packet retains a `memory' of the incident wave packet by virtue of its underlying spatio-temporal structure. In paper (I) \cite{Yessenov21JOSAA} we formulated a law of refraction for `baseband' ST wave packets \cite{Yessenov19PRA} at normal incidence on a planar interface, which predicts surprising phenomena stemming from this `memory' effect.

Here, we verify experimentally the refractive phenomena that occur when baseband ST wave packets are incident normally at a planar interface between two non-dispersive, homogeneous, isotropic dielectrics. The first phenomenon is \textit{group-velocity invariance}, which refers to the existence of an ST wave packet for any given pair of media that retains its group velocity after traversing a planar interface separating them regardless of their index contrast. The second is \textit{anomalous refraction}, which denotes a regime in which the group velocity of a transmitted ST wave packet increases in a higher-index medium in contrast to the usual expectation. The third phenomenon is \textit{group-velocity inversion}, which refers to the existence for any pair of media of an ST wave packet that retains the magnitude of its group velocity while switching sign across the interface. As a corollary, we also confirm \textit{group-delay cancellation} whereby the group delay incurred vanishes upon traversing equal lengths of two media when the condition for group-velocity inversion is satisfied. In this paper, we verify experimentally in detail these predicted phenomena in a variety of optical materials: MgF$_2$, fused silica, BK7 glass, and sapphire, with refractive indices of $n\!=\!1.38$, 1.45, 1.51, and 1.76, respectively, at a wavelength of $\lambda\!\sim\!800$~nm \cite{Bhaduri19Optica}. 

Such surprising phenomena point to non-trivial refraction-driven dynamics exhibited by the spectral support domain of ST wave packets. For conventional wave packets, a transition to a higher-index medium is accompanied by a stretching of the spatio-temporal spectral support domain because of the `inflation' of the light-cone surface with refractive index. Such a stretching reduces the velocity in the higher-index medium. Anomalous refraction of ST wave packets therefore indicates that light-cone inflation can sometimes be accompanied -- surprisingly -- by a spatio-temporal spectral \textit{shrinkage} rather than stretching, which results in an \textit{increase} in group velocity, and hence anomalous refraction. We describe these spectral dynamics on the light-cone and uncover a hitherto unnoticed geometrical effect that undergirds the anomalous refraction of ST wave packets. As described in paper (I), these phenomena occur upon the refraction of baseband ST wave packets \cite{Yessenov19PRA}, and do \textit{not} occur for other kinds -- such as X-waves \cite{Saari97PRL} and sideband ST wave packets (e.g., focus-wave modes \cite{Brittingham83JAP}), whose refraction has been studied theoretically \cite{Hillion93Optik,Donnelly97IEEE,Hillion98JO,Hillion99JOA,Attiya01PER,Shaarawi01PER,Salem12JOSAA}. Furthermore these phenomena do \textit{not} occur with conventional pulsed beams whose spatial and temporal degrees of freedom are separable \cite{Kondakci19OL}.

This paper is structured as follows. First, we briefly review the law of refraction for baseband ST wave packets at normal incidence. We then describe the experimental arrangement and measurement procedure we utilize to verify the predicted refractive phenomena at normal incidence. Next, we confirm group-velocity invariance, the transition from normal to anomalous refraction, group-velocity inversion, and group-delay cancellation in the optical materials listed above. Finally, we provide a geometric representation of the dynamics of the spectral support domain for ST wave packets on the surface of the light-cone, which elucidates the origin of the counter-intuitive transition from normal to anomalous refraction, and the existence of group-velocity-invariant ST wave packets.

\section{Refraction of baseband space-time wave packets at normal incidence}

We briefly review the theoretical basis for the refraction of ST wave packets at normal incidence upon a planar interface between two non-dispersive, homogeneous, isotropic dielectrics of refractive indices $n_{1}$ and $n_{2}$. The spectral support domain of a ST wave packet in a medium of index $n$ lies at the intersection of the light-cone $k_{x}^{2}+k_{z}^{2}\!=\!n^{2}(\tfrac{\omega}{c})^{2}$ with a tilted plane $\mathcal{P}_{\mathrm{B}}(\theta)$ defined by the equation $\omega\!=\!\omega_{\mathrm{o}}+(k_{z}-nk_{\mathrm{o}})c\tan{\theta}$. Here $k_{x}$ is the transverse component of the wave vector along $x$ (or the spatial frequency), $k_{z}$ is its axial component, $\omega$ is the temporal (angular) frequency, $c$ is the speed of light in vacuum, $\omega_{\mathrm{o}}$ is a fixed frequency, $k_{\mathrm{o}}\!=\!\omega_{\mathrm{o}}/c$ is the corresponding free-space wave number, and the field is assumed uniform along $y$ ($k_{y}\!=\!0$). The plane $\mathcal{P}_{\mathrm{B}}(\theta)$ is parallel to the $k_{x}$-axis and is tilted with respect to the $k_{z}$-axis by an angle $0\!<\!\theta\!<\!180^{\circ}$, denoted the spectral tilt angle. This optical field represents a propagation-invariant wave packet traveling in the medium at a group velocity $\widetilde{v}\!=\!c\tan{\theta}\!=\!c/\widetilde{n}$, where $\widetilde{n}\!=\!\cot{\theta}$ is the group index \cite{Kondakci17NP,Yessenov19PRA}, with negative-valued group velocities occurring when $\theta\!>\!90^{\circ}$. The `luminal' velocity in the medium is $\widetilde{v}\!=\!c/n$ corresponding to $\widetilde{n}\!=\!n$. It is critical to note that the group velocity can be tuned by changing the field structure to realize a different spectral tilt angle $\theta$. This is \textit{not} related to any chromatic dispersion in the medium, which we assume to be non-dispersive \cite{Bhaduri19Optica,Bhaduri20NP}.

\begin{figure}[t!]
\centering
\includegraphics[width=8cm]{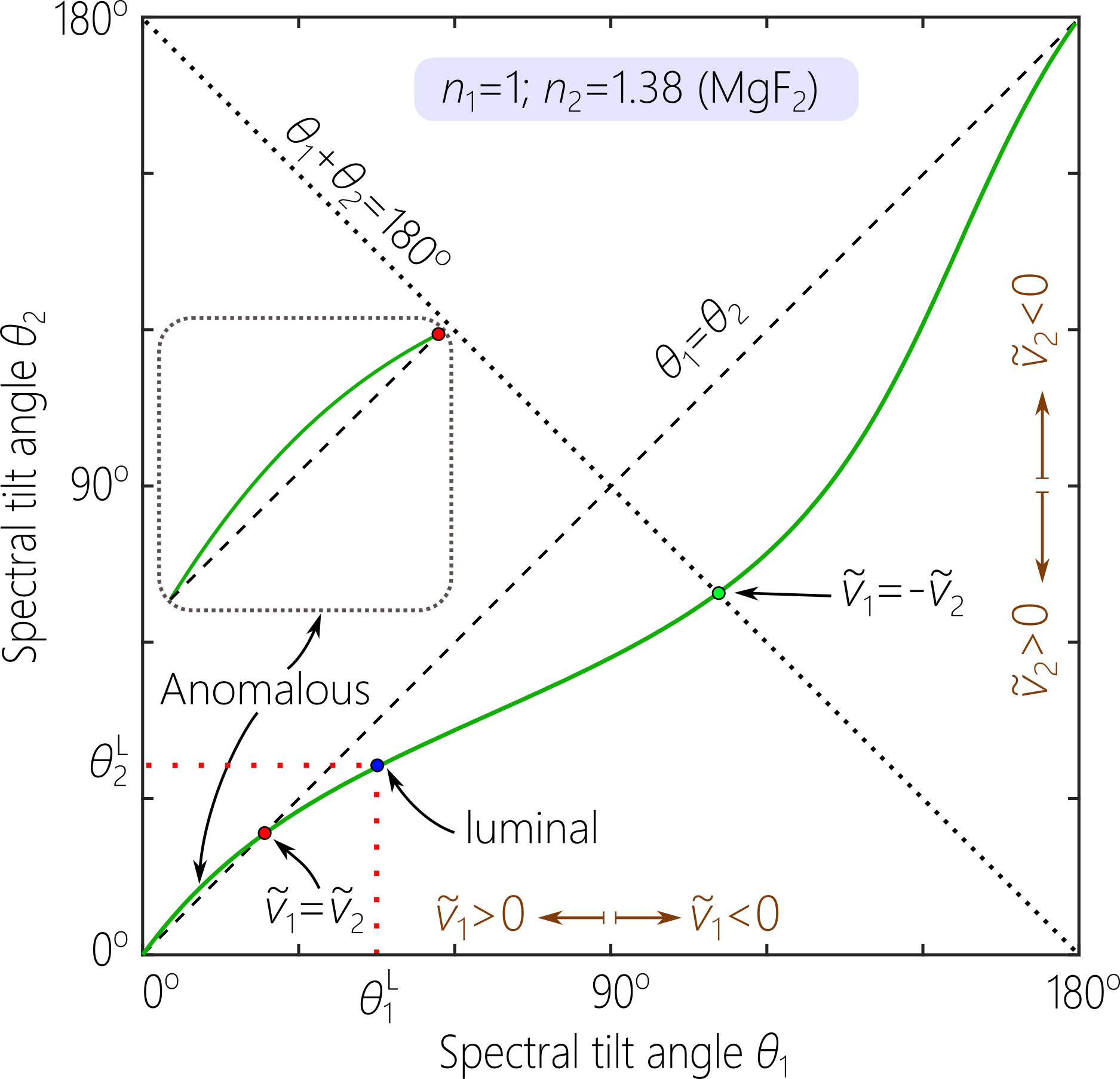}
\caption{Law of refraction for ST wave packets at normal incidence (solid curve) from a medium of refractive index $n_{1}$ onto one of index $n_{2}$, plotted in terms of the spectral tilt angles $\theta_{1}$ and $\theta_{2}$ of the incident and transmitted wave packets, respectively. The curve is plotted for $n_{1}\!=\!1$ (free space) and $n_{2}\!=\!1.38$ (MgF$_2$), but the generic features of the plot apply to all material combinations. The intersection of the curve with the diagonal $\theta_{1}\!=\!\theta_{2}$ corresponds to group-velocity invariance $\widetilde{v}_{1}\!=\!\widetilde{v}_{2}$, the intersection with the anti-diagonal $\theta_{1}+\theta_{2}\!=\!180^{\circ}$ corresponds to group-velocity inversion $\widetilde{v}_{2}\!=\!-\widetilde{v}_{1}$, and the luminal condition is the point where $\cot{\theta_{1}^{\mathrm{L}}}\!=\!n_{1}$ and $\cot{\theta_{2}^{\mathrm{L}}}\!=\!n_{2}$. The inset highlights the region associated with anomalous refraction.}
\label{Fig:LawOfRefraction}
\end{figure}

\begin{figure*}[t!]
  \begin{center}
  \includegraphics[width=12.5cm]{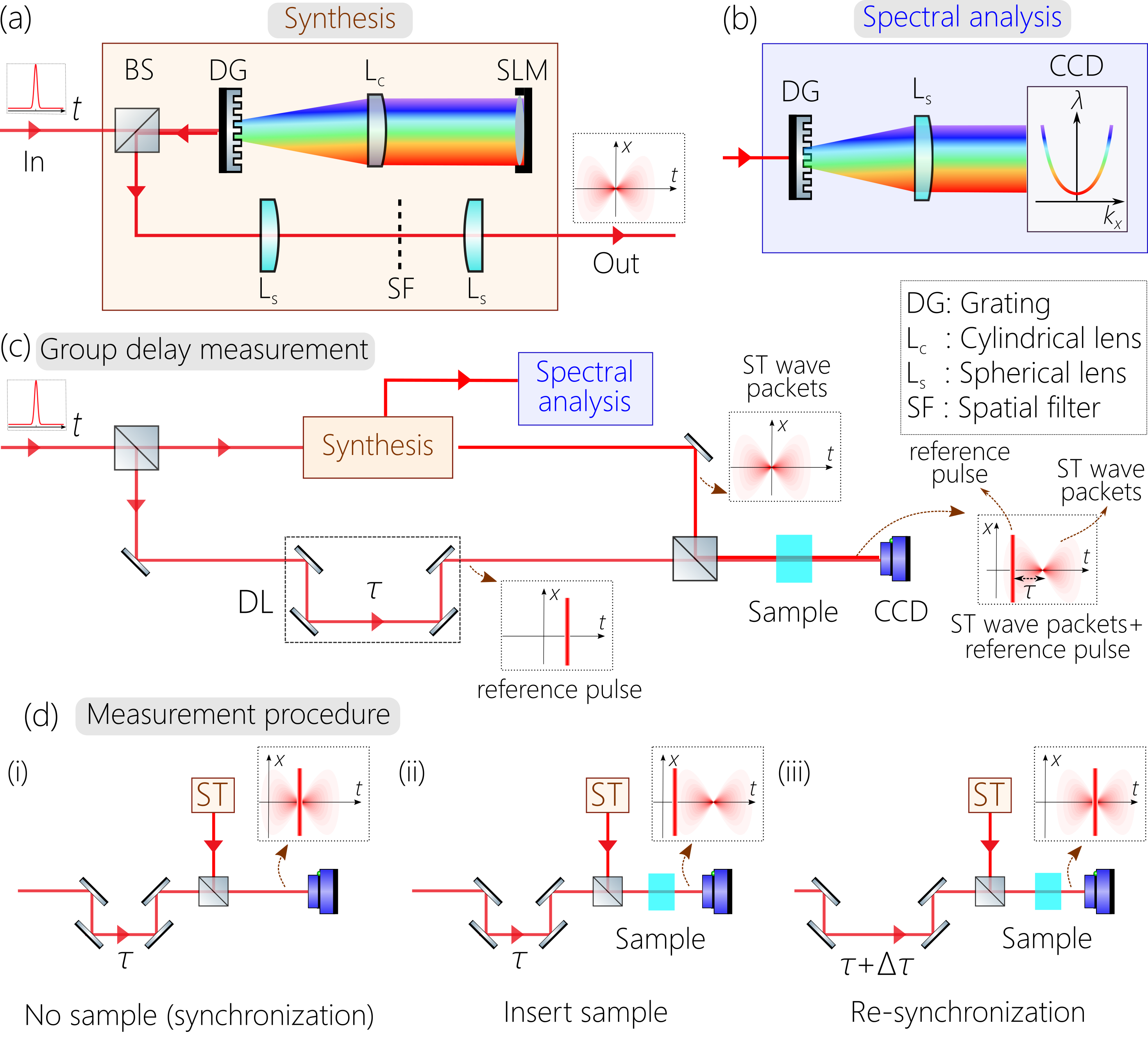}
  \end{center}
  \caption{Measurement strategy for studying the refraction of ST wave packets. (a) Synthesis starts from an ultrafast plane-wave pulse (left) and produces an ST wave packet (right) of controllable group velocity. A grating spreads the pulse spectrum, a cylindrical lens collimates the wave front, and a SLM modulates the wave-front phase to realize a prescribed spectral tilt angle $\theta$. (b) A spatio-temporal spectral analysis setup extracts the spectral projection onto the $(k_{x},\lambda)$-plane. (c) The configuration for measuring the group delay accrued by a ST wave packet after traversing a sample layer. The arrangements for synthesis from (a) and that for spectral analysis from (b) are included in one arm of a Mach-Zehnder interferometer. The initial laser pulse traverses the reference arm. (d) The measurement procedure to estimate the group delay of the ST wave packets traversing a sample; see text for details.}
  \label{Fig:Setup}
\end{figure*}

The spectral support domain takes the form of a conic section: a circle at $\theta\!=\!0^{\circ}$ corresponding to a monochromatic beam; a straight line when $\tan{\theta}\!=\!\tfrac{1}{n}$, corresponding to a luminal plane-wave pulse; a parabola when $\tan{\theta}\!=\!-\tfrac{1}{n}$ corresponding to a negative-velocity luminal ST wave packet; a hyperbola when $-\tfrac{1}{n}\!<\!\tan{\theta}\!<\!\tfrac{1}{n}$; and an ellipse otherwise \cite{Yessenov19PRA}. In all these cases (except when $\theta\!=\!0$ or $\tan{\theta}\!=\!\tfrac{1}{n}$), the conic section can be approximated in the narrowband paraxial regime by a parabola in the vicinity of $k_{x}\!=\!0$:
\begin{equation}\label{Eq:ParaxialParabola}
\frac{k_{x}^{2}}{2k_{\mathrm{o}}^{2}}=n(n-\widetilde{n})\frac{\omega-\omega_{\mathrm{o}}}{\omega_{\mathrm{o}}}.
\end{equation}

The central theoretical insight in paper (I) is that the quantity $n(n-\widetilde{n})$ is a \textit{refractive invariant} for baseband ST wave packets, which is the curvature of the spectral support domain in the vicinity of $k_{x}\!=\!0$ when projected onto the $(k_{x},\tfrac{\omega}{c})$-plane. This quantity is invariant across planar interfaces, leading to the following law of refraction:
\begin{equation}\label{Eq:LawOfRefraction}
n_{1}(n_{1}-\widetilde{n}_{1})=n_{2}(n_{2}-\widetilde{n}_{2});
\end{equation}
where $\widetilde{n}_{1}\!=\!\cot{\theta_{1}}$ and $\widetilde{n}_{2}\!=\!\cot{\theta_{2}}$, and $\theta_{1}$ and $\theta_{2}$ are the spectral tilt angles for the ST wave packet in the two media. This relationship is plotted  in Fig.~\ref{Fig:LawOfRefraction} in terms of $\theta_{1}$ and $\theta_{2}$. The group velocity of the transmitted ST wave packet $\widetilde{v}_{2}$ does not depend solely on $n_{2}$. Instead $\widetilde{v}_{2}$ depends on both $n_{1}$ and $n_{2}$, in addition to the group velocity of the incident wave packet $\widetilde{v}_{1}$, which is a manifestation of the above-described `memory' effect. Here, we explore experimentally several surprising consequences of this law of refraction.

\section{Experimental strategy}

\subsection{Measurement setup}

Our experimental arrangement is depicted in Fig.~\ref{Fig:Setup} and comprises several subsystems. The first subsystem is that for \textit{synthesizing} ST wave packets in free space with prescribed spectral tilt angle $\theta$ [Fig.~\ref{Fig:Setup}(a)], which was developed in our previous work \cite{Kondakci17NP,Yessenov19OPN}. The source is a Ti:sapphire laser (Spectra Physics, Tsunami) producing pulses at a central wavelength $\sim\!800$~nm and bandwidth $\sim\!10$~nm (pulsewidth $\sim\!100$~fs). A two-dimensional pulse shaper utilizing a spatial light modulator (SLM; Hamamatsu X10468-02) imprints a phase pattern on the spectrally resolved wave front to assign to each wavelength $\lambda$ a spatial frequency $\pm k_{x}$ according to the relationship in Eq.~\ref{Eq:ParaxialParabola}, thereby allowing for the realization of a ST wave packet with prescribed $\theta$. This arrangement spectrally filters the field to a bandwidth of $\Delta\lambda\!\approx\!0.25$~nm, thereby resulting in synthesized ST wave packets whose on-axis pulsewidth is $\sim\!9$~ps.

To characterize the ST wave packet, two other subsystems are implemented. First, to measure the spatio-temporal spectrum we make use of the configuration in Fig.~\ref{Fig:Setup}(b) that comprises a grating to resolve the wavelengths and a lens in a $2f$ configuration to map the spatial frequencies to points in the focal plane. This spatio-temporal Fourier transform reveals the structured spectrum of the ST wave packets, from which we can confirm that the prescribed $\theta$ has been realized \cite{Kondakci17NP,Kondakci18OE}. Second, to trace the spatio-temporal intensity profile of the ST wave packet $I(x,z;\tau)$ at a fixed axial plane $z$, we interfere the synthesized ST wave packet with the original Ti:sapphire laser pulse after placing the synthesis arrangement in one arm of a Mach-Zehnder interferometer and inserting an optical delay $\tau$ in the reference arm \cite{Kondakci19NC,Bhaduri19Optica}. The superposition of the ST wave packet and the plane-wave reference pulse are directed to an axially translatable CCD camera; see Fig.~\ref{Fig:Setup}(c). When the two wave packets overlap in space and time, spatially resolved interference fringes are observed, whose visibility enables reconstructing the intensity of the ST wave packet envelope $I(x,z;\tau)$. Finally, the axial evolution of the time-averaged intensity (or energy) of the ST wave packet $I(x,z)\!=\!\int\!d\tau I(x,z;\tau)$ is captured with the axially scanned CCD camera after blocking the reference arm. 

The materials we use in our measurements are MgF$_2$ with $n\!=\!1.38$ (Thorlabs, WG61050) \cite{Dodge84AO}, UV fused silica (henceforth silica for brevity) with $n\!=\!1.45$ (Thorlabs, BSF20-B) \cite{Maltison65JOSA}, BK7 glass with $n\!=\!1.51$ (Thorlabs, WG12012-B) \cite{Jedamzik14ProcSPIE}, and sapphire with $n\!=\!1.76$ (Thorlabs, WG31050) \cite{Malitson62JOSA}; the values for $n$ are measured at a wavelength of $\lambda\!\sim\!800$~nm \cite{Bhaduri19Optica}. The lengths of the samples are all $L\!=\!5$~mm, except for BK7 where $L\!=\!12$~mm.

\subsection{Measurement procedure}

Our goal is to measure the group delay incurred by a ST wave packet synthesized in free space (spectral tilt angle $\theta_{0}$ and group index $\widetilde{n}_{0}\!=\!\cot{\theta_{0}}$) after traversing a layer of thickness $L_{1}$ and refractive index $n_{1}$ at normal incidence. From this delay we can estimate the group velocity of the refracted ST wave packet in the medium. We first synchronize the ST wave packet with the reference pulse in absence of the layer by overlapping them in space and time [panel (i) in Fig.~\ref{Fig:Setup}(d)]. We then place the layer in the common path of the ST wave packet and the reference pulse while maintaining the position of the detector fixed [panel (ii) in Fig.~\ref{Fig:Setup}(d)]. The ST wave packet incurs an additional group delay $\tau_{\mathrm{ST}}\!=\!L_{1}(\widetilde{n}_{1}-\widetilde{n}_{0})/c$ in presence of the layer with respect to its absence, whereas the reference pulse incurs $\tau_{\mathrm{R}}\!=\!L_{1}(n_{1}-n_{0})/c$, where $n_{0}$ is the index of the ambient medium. Because of the group-delay difference $\tau_{\mathrm{ST}}-\tau_{\mathrm{R}}\!=\!L_{1}\{(\widetilde{n}_{1}-n_{1})-(\widetilde{n}_{0}-n_{0})\}/c$, the visibility of the spatially resolved interference fringes drops and may vanish altogether (if the relative group delay is larger than the coherence time). However, the interference can be retrieved by adding a path length $\Delta\ell_{1}$ in the reference arm [panel (iii) in Fig.~\ref{Fig:Setup}(d)], corresponding to a delay $\Delta\tau_{1}\!=\!\Delta\ell_{1}n_{0}/c$. When $\Delta\tau_{1}\!=\!\tau_{\mathrm{ST}}-\tau_{\mathrm{R}}$, the ST wave packet and the reference pulse overlap once again in space and time. From the relationship $L_{1}\{(\widetilde{n}_{1}-n_{1})-(\widetilde{n}_{0}-n_{0})\}\!=\!\Delta\ell_{1}n_{0}$ and knowledge of the refractive indices $n_{0}$ and $n_{1}$, the thickness $L_{1}$, the additional delay $\Delta\ell_{1}$, and the group index of the incident wave packet $\widetilde{n}_{0}$, we can determine the group index $\widetilde{n}_{1}$ in the sample and hence $\theta_{1}$. By sweeping $\theta_{0}$ through sculpting the phase distribution imparted by the SLM in Fig.~\ref{Fig:Setup}(a) \cite{Yessenov19PRA}, we can determine the corresponding $\theta_{1}$ in the sample and thus trace the law of refraction in Eq.~\ref{Eq:LawOfRefraction}, as plotted in Fig.~\ref{Fig:LawOfRefraction} and confirmed experimentally in Ref.~\cite{Bhaduri20NP}.

In the case of a bilayer of materials of indices $n_{1}$ and $n_{2}$ and corresponding thicknesses $L_{1}$ and $L_{2}$ ($L\!=\!L_{1}+L_{2}$), the above strategy can be applied to each layer separately leading to two conditions $L_{1}\{(\widetilde{n}_{1}-n_{1})-(\widetilde{n}_{0}-n_{0})\}\!=\!\Delta\ell_{1}n_{0}$ and $L_{2}\{(\widetilde{n}_{2}-n_{2})-(\widetilde{n}_{0}-n_{0})\}\!=\!\Delta\ell_{2}n_{0}$. When the ST wave packet is incident on the bilayer, we have $L_{1}(\widetilde{n}_{1}-n_{1})+L_{2}(\widetilde{n}_{2}-n_{2})-L(\widetilde{n}_{0}-n_{0})\!=\!\Delta\ell n_{0}$, with $\Delta\ell\!=\!\Delta\ell_{1}+\Delta\ell_{2}$. From these measurements we can extract the relationship between $\theta_{1}$ for the ST wave packet in the first medium and $\theta_{2}$ in the second.

\begin{figure}[t!]
\centering
\includegraphics[width=8.6cm]{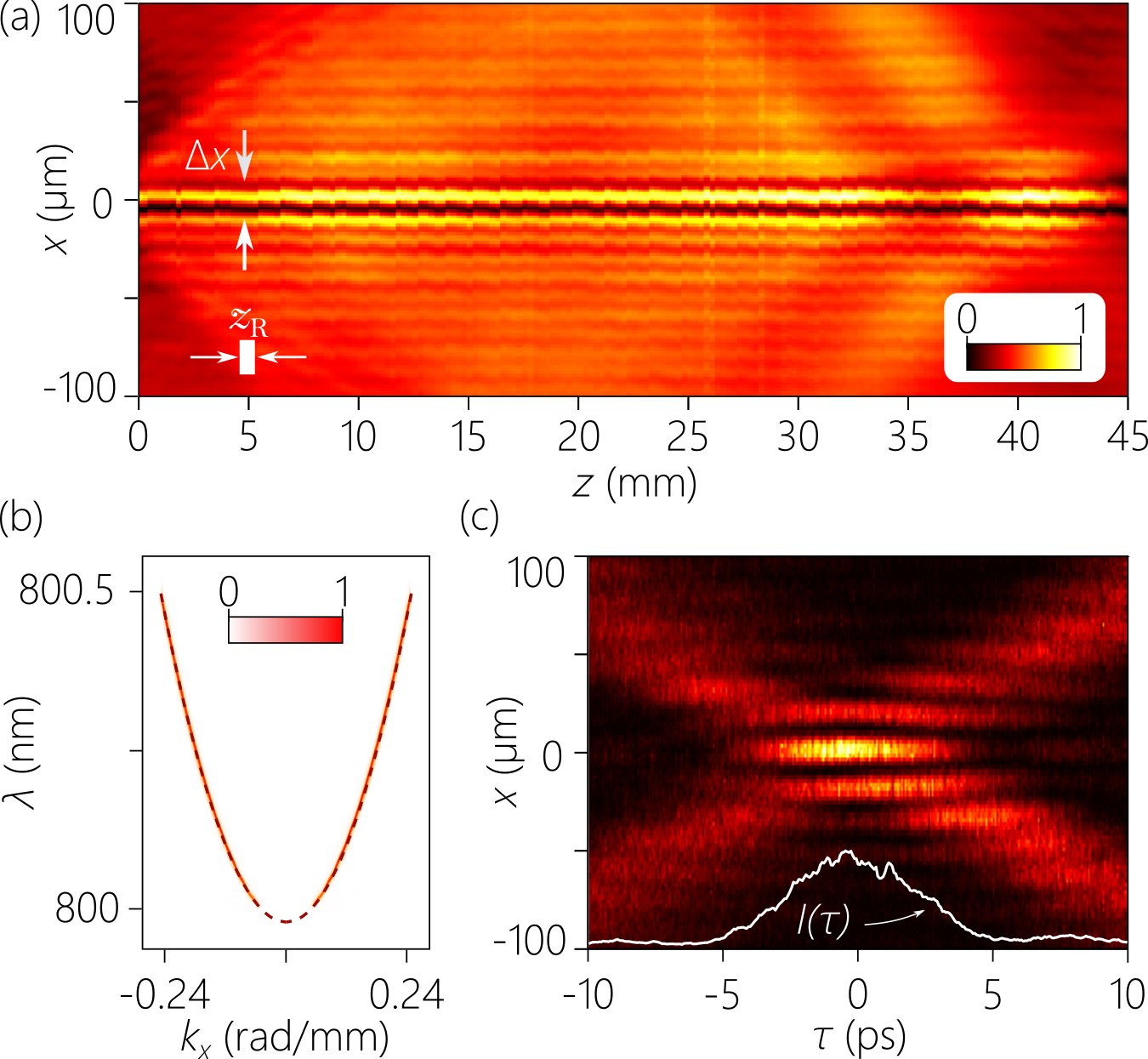}
\caption{(a) Measured time-averaged intensity $I(x,z)$ along the propagation axis for a baseband ST wave packet with $\theta\!=\!30^{\circ}$ in free space. Note the different length scales along $x$ and $z$. The white rectangle corresponds to the Rayleigh range $z_{\mathrm{R}}$ of a Gaussian beam having the same beam waist $\Delta x$ as the width of the main lobe of the ST wave packet. (b) Measured spatio-temporal spectral intensity of the wave packet in $(k_{x},\lambda)$-space. The dashed curve is the theoretical expectation for $\theta\!=\!30^{\circ}$ from Eq.~\ref{Eq:ParaxialParabola}. (c) Measured spatio-temporal intensity profile $I(x,z;\tau)$ of the ST wave packet at $z\!=\!0$. The white curve is the on-axis pulse profile $I(\tau)\!=\!I(0,0;\tau)$.}
\label{Fig:Measurements}
\end{figure}

We plot in Fig.~\ref{Fig:Measurements} an example characterizing a ST wave packet having a spectral tilt angle of $\theta\!=\!30^{\circ}$ in free space: the time-averaged intensity showing diffraction-free propagation [Fig.~\ref{Fig:Measurements}(a)]; the spatio-temporal spectrum in the $(k_{x},\lambda)$-plane [Fig.~\ref{Fig:Measurements}(b)]; and the spatio-temporal profile $I(x,z;\tau)$ at $z\!=\!0$ [Fig.~\ref{Fig:Measurements}(c)].

\section{Group-velocity invariance}

\begin{figure}[t!]
\centering
\includegraphics[width=7cm]{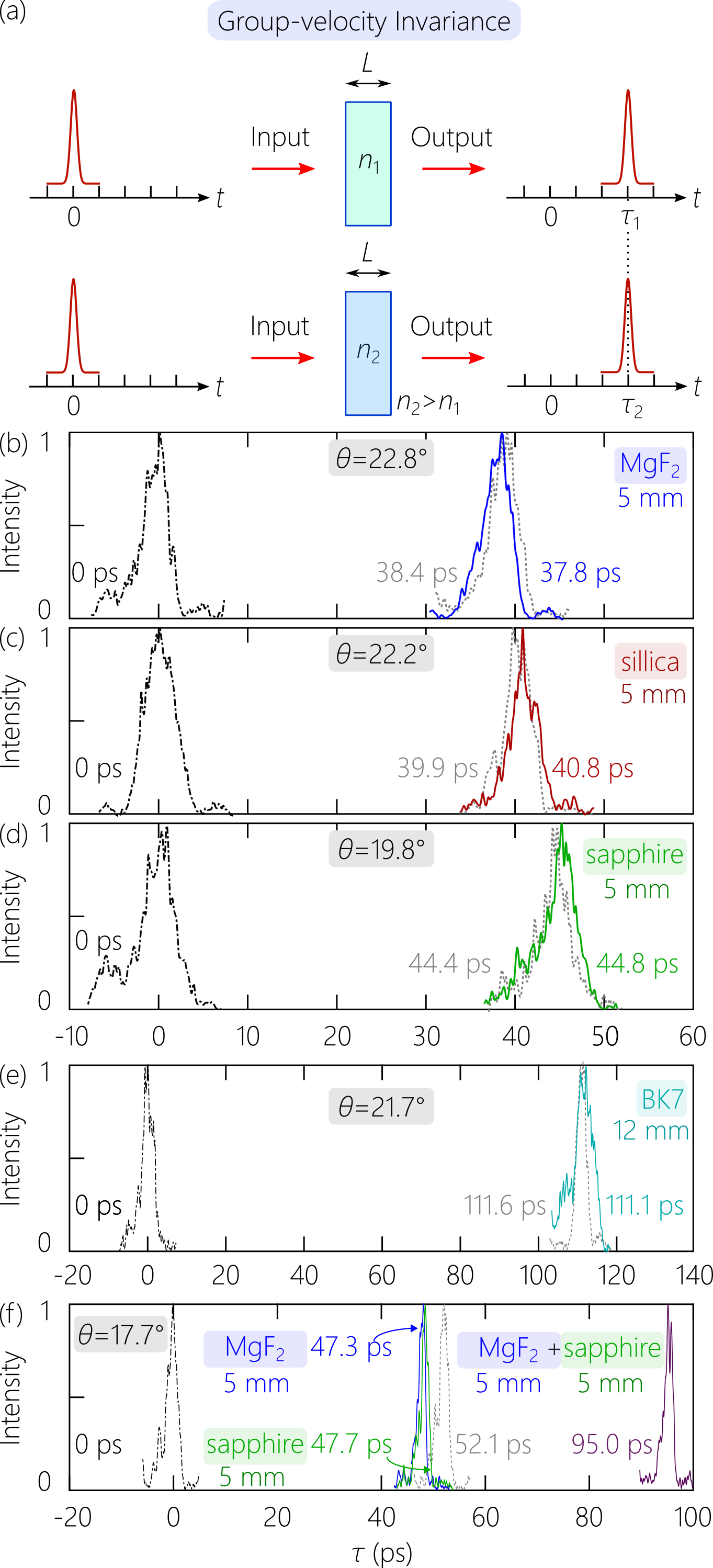}
\caption{\small{Concept of group-velocity invariance for ST wave packets. (a) Schematic of the concept. (b-e) Plots of the temporal profile of the incident wave packet (dashed-dotted curve on the left), and those of the wave packets after traversing a sample layer of thickness $L$ (solid curve) and traversing an equal length of free space (dotted curve). We indicate the measured group delays for the two wave packets, which overlap in all cases. In (b), the material is MgF$_2$, with $\widetilde{n}_{1}\!=\!2.38$ ($\theta_{1}\!=\!22.8^{\circ}$); (c) silica, with $\widetilde{n}_{1}\!=\!2.45$ ($\theta_{1}\!=\!22.2^{\circ}$); (d) sapphire, with $\widetilde{n}_{1}\!=\!2.76$ ($\theta_{1}\!=\!19.8^{\circ}$); and (e) BK7, with $\widetilde{n}_{1}\!=\!2.51$ ($\theta_{1}\!=\!21.7^{\circ}$). In (b-d) $L\!=\!5$~mm, and in (e) $L\!=\!12$~mm. (f) Group-velocity invariance in a bilayer of MgF$_2$ and sapphire; $\widetilde{n}_{0}\!=\!3.42$ in free space ($\theta_{0}\!\approx\!17.7^{\circ}$). We plot the temporal profiles after each layer, 5~mm of free space (dotted curve), and in a bilayer.}}
\label{Fig:GroupVelocityInvarianceFromFreeSpace}
\end{figure}

A unique feature of ST wave packets is that there always exists one that traverses an interface with no change in its group velocity $\widetilde{v}_{1}\!=\!\widetilde{v}_{2}$, regardless of the index contrast. We refer to this condition as \textit{group-velocity invariance}. Substituting $\widetilde{n}_{1}\!=\!\widetilde{n}_{2}\!=\!\widetilde{n}_{\mathrm{th}}$ in Eq.~\ref{Eq:LawOfRefraction} yields the solution:
\begin{equation}
\widetilde{n}_{\mathrm{th}}\!=\!n_{1}+n_{2}.
\end{equation}
That is, a ST wave packet with $\widetilde{n}_{1}\!=\!\widetilde{n}_{\mathrm{th}}$ traverses the interface such that $\widetilde{n}_{2}\!=\!\widetilde{n}_{\mathrm{th}}\!=\!\widetilde{n}_{1}$, thereby maintaining the group velocity fixed. Because $\widetilde{n}_{\mathrm{th}}\!>\!n_{1},n_{2}$, this wave packet is subluminal in both media. For such a ST wave packet, the group delay accrued over a layer of length $L$ and index $n_{1}$ is equal to that in a layer of the same thickness $L$ and index $n_{2}$ [Fig.~\ref{Fig:GroupVelocityInvarianceFromFreeSpace}(a)].

To demonstrate group-velocity invariance, we take the two media to be free space ($n_{1}\!=\!1$) and a dielectric (MgF$_2$, fused silica, BK7, or sapphire; $n_{2}\!=\!n$), whereupon $\widetilde{n}_{\mathrm{th}}\!=\!1+n$. When such a ST wave packet is incident from free space onto a layer of thickness $L$ and index $n$, the group delay incurred in the layer is equal to that across a distance $L$ in free space. Group-delay measurements are plotted in Fig.~\ref{Fig:GroupVelocityInvarianceFromFreeSpace}(b-e) for the four materials when the condition for group-velocity invariance is satisfied. In each case we compare the group delay incurred upon traversing the layer with the group delay traversing the \textit{same length} of free space. We synthesize for each medium a different ST wave packet whose group index in free space is $\widetilde{n}_{\mathrm{th}}\!=\!1+n$, whereupon $\theta_{1}\!=\!22.8^{\circ}$ for MgF$_2$, $\theta_{1}\!=\!22.2^{\circ}$ for silica, $\theta_{1}\!=\!19.8^{\circ}$ for sapphire, and $\theta_{1}\!=\!21.7^{\circ}$ for BK7. For each material we plot the temporal profile of the ST wave packet at the entrance to the layer, and then the profiles after traversing the material layer and after traversing an equal length of free space. In all cases, the two profiles for the output wave packets overlap, thus indicating that the group delays are approximately equal.

As further confirmation of group-velocity invariance, we carried out measurements for a bilayer comprising a pair of materials; here MgF$_2$ and sapphire ($L\!=\!5$~mm each). The threshold index is $\widetilde{n}_{\mathrm{th}}\!=\!n_{1}+n_{2}$ for the ST wave packet within both media. The group index of the ST wave packet incident from free space necessary to realize this condition is $\widetilde{n}_{0}\!=\!1+n_{1}n_{2}$. The group delays accrued by this ST wave packet in each layer are approximately equal ($\approx\!47.3$~ps and 47.7~ps). In contrast, the delay after propagation in free space for an equal distance is $\approx\!52.1$~ps, which is no longer equal to the delay in either layer. The group delay accrued after traversing a bilayer of MgF$_2$ and sapphire them is found to be approximately double that in each layer separately; see Fig. \ref{Fig:GroupVelocityInvarianceFromFreeSpace}(f).

\begin{figure}[t!]
\centering
\includegraphics[width=8cm]{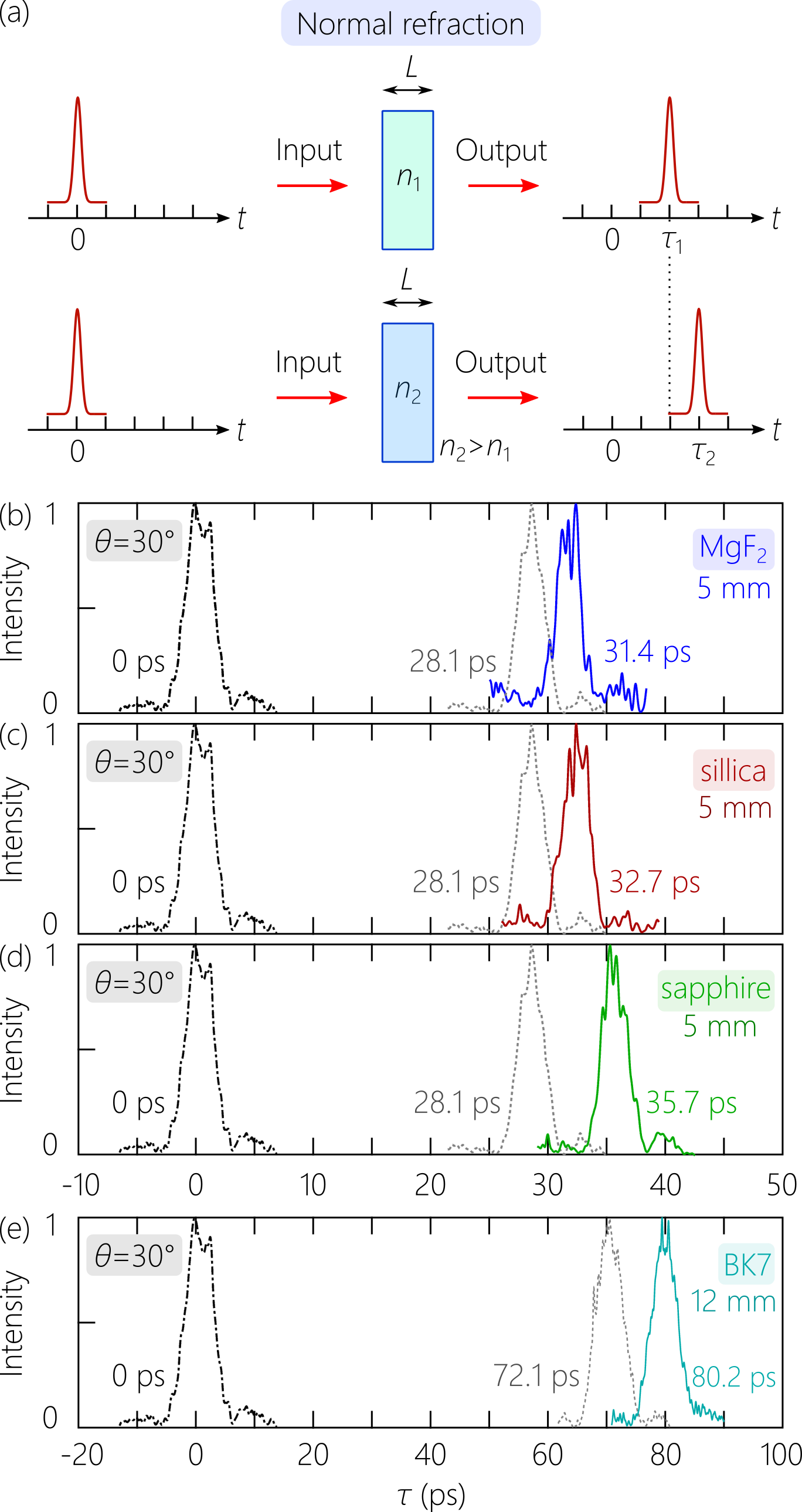}
\caption{(a) Concept of normal refraction for ST wave packets. (b-e) Plots of the temporal profile of the incident wave packet (dashed-dotted curve on the left), and after traversing a sample layer of thickness $L$ (solid curves on the right) and after traversing an equal-length of free space (dotted curves): (b) MgF$_2$, (c) silica, (d) sapphire, and (e) BK7. The same input ST wave packet is used throughout with $\widetilde{n}_{1}\!=\!1.73$ ($\theta_{1}\!=\!30^{\circ}$; $\theta_{1}\!>\!\theta_{\mathrm{th}}$ in all materials). We provide the measured group delays for the two wave packets, with the wave packet always emerging from the material layer later than from free space. In (b-d) $L\!=\!5$~mm and in (e) $L\!=\!12$~mm.}
\label{Fig:NormalRefractionFromFreeSpace}
\end{figure}

\begin{figure}[t!]
\centering
\includegraphics[width=8cm]{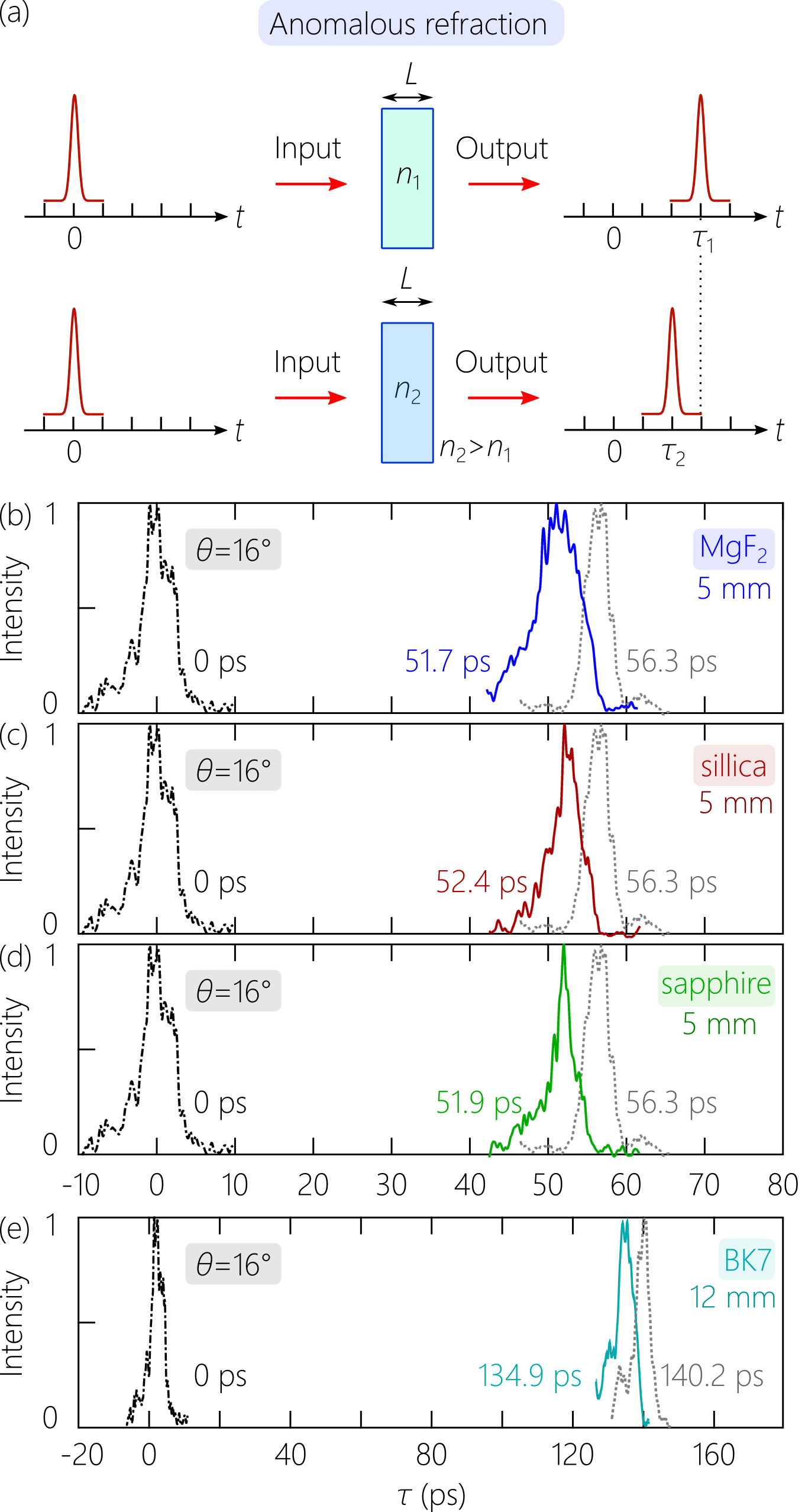}
\caption{(a) Concept of anomalous refraction for ST wave packets. (b-e) Plots of the temporal profile of the incident wave packet (dashed-dotted curve on the left), and after traversing a sample layer of thickness $L$ (solid curves on the right) and after traversing an equal-length of free space (dotted curves): (b) MgF$_2$, (c) silica, (d) sapphire, and (e) BK7. The same input ST wave packet is used throughout with $\widetilde{n}_{1}\!=\!3.49$ ($\theta_{1}\!=\!16^{\circ}$; $\theta_{1}\!<\!\theta_{\mathrm{th}}$ in all materials). We indicate the measured group delays for the two wave packets, with the wave packet always emerging from the material layer earlier than from free space. In (b-d) $L\!=\!5$~mm and in (e) $L\!=\!12$~mm.}
\label{Fig:AnomalousRefractionFromFreeSpace}
\end{figure}

\section{Normal and anomalous refraction}

A ST wave packet of group index $\widetilde{n}_{1}\!=\!\widetilde{n}_{\mathrm{th}}\!=\!n_{1}+n_{2}$ traverses the interface at normal incidence while retaining its group velocity. This particular group index represents a threshold separating two distinct regimes: normal and anomalous refraction [Fig.~\ref{Fig:LawOfRefraction}]. In the \textit{normal} refraction regime $\widetilde{n}_{1}\!<\!\widetilde{n}_{\mathrm{th}}$, the group velocity of the wave packet transmitted from low-index to high-index decreases as usual for conventional wave packets; that is $\widetilde{n}_{2}\!>\!\widetilde{n}_{1}$ ($\widetilde{v}_{2}\!<\!\widetilde{v}_{1}$) when $n_{2}\!>\!n_{1}$ [Fig.~\ref{Fig:NormalRefractionFromFreeSpace}(a)]. This regime encompasses all superluminal ST wave packets $\widetilde{n}_{1}\!<\!n_{1}$, in addition to a portion of the subluminal regime $n_{1}\!<\!\widetilde{n}_{1}\!<\!n_{\mathrm{th}}$. On the other hand, in the \textit{anomalous} refraction regime $\widetilde{n}_{1}\!>\!\widetilde{n}_{\mathrm{th}}$ (thus comprising only part of the subluminal regime), the group velocity of the wave packet transmitted from low-index to high-index \textit{increases} against expectations; that is, $\widetilde{n}_{2}\!<\!\widetilde{n}_{1}$ ($\widetilde{v}_{2}\!>\!\widetilde{v}_{1}$) despite $n_{2}\!>\!n_{1}$ [Fig.~\ref{Fig:AnomalousRefractionFromFreeSpace}(a) and Fig.~\ref{Fig:LawOfRefraction}, inset]. To the best of our knowledge, this is the unique optical-field configuration exhibiting anomalous refraction amongst \textit{all} optical fields in non-dispersive media, including X-waves and sideband ST wave packets.

To confirm the predictions of \textit{normal} refraction when $\widetilde{n}\!<\!\widetilde{n}_{\mathrm{th}}$, we carried out experiments with the four dielectrics used above, and make use of an incident ST wave packet with $\theta_{1}\!=\!30^{\circ}$, which is larger than $\theta_{\mathrm{th}}$ for all the materials. Such a wave packet undergoes a group delay of $\approx\!28.1$~ps and $\approx\!72.1$~ps in 5~mm and 12~mm of free space, respectively. In Fig.~\ref{Fig:NormalRefractionFromFreeSpace}(b-e) we plot the wave packet after traversing a dielectric layer and an equal length of free space. In all cases, the group delay incurred after traversing the dielectric layer is larger than that over the same distance in free space, thus confirming normal refraction throughout.

To confirm the predictions of \textit{anomalous} refraction when $\widetilde{n}_{1}\!>\!\widetilde{n}_{\mathrm{th}}$, we repeated the measurements with the four dielectrics using an incident wave packet whose spectral tilt angle is $\theta_{1}\!=\!16^{\circ}$ in free space, which is smaller than $\theta_{\mathrm{th}}$ in all the materials. Such a wave packet accrues a group delay of 56.3~ps and 140.2~ps in 5~mm and 12~mm of free space, respectively. The measurement results are plotted for this wave packet in Fig.~\ref{Fig:AnomalousRefractionFromFreeSpace}(b-e) after traversing a sample layer and equal length of free space. In all cases, the group delay incurred after traversing the sample layer is \textit{smaller} than that over an equal length of free space. This provides unambiguous confirmation that we are in the anomalous refraction regime throughout.

\section{Group-velocity inversion and group-delay cancellation}

\begin{figure}[t!]
\centering
\includegraphics[width=8cm]{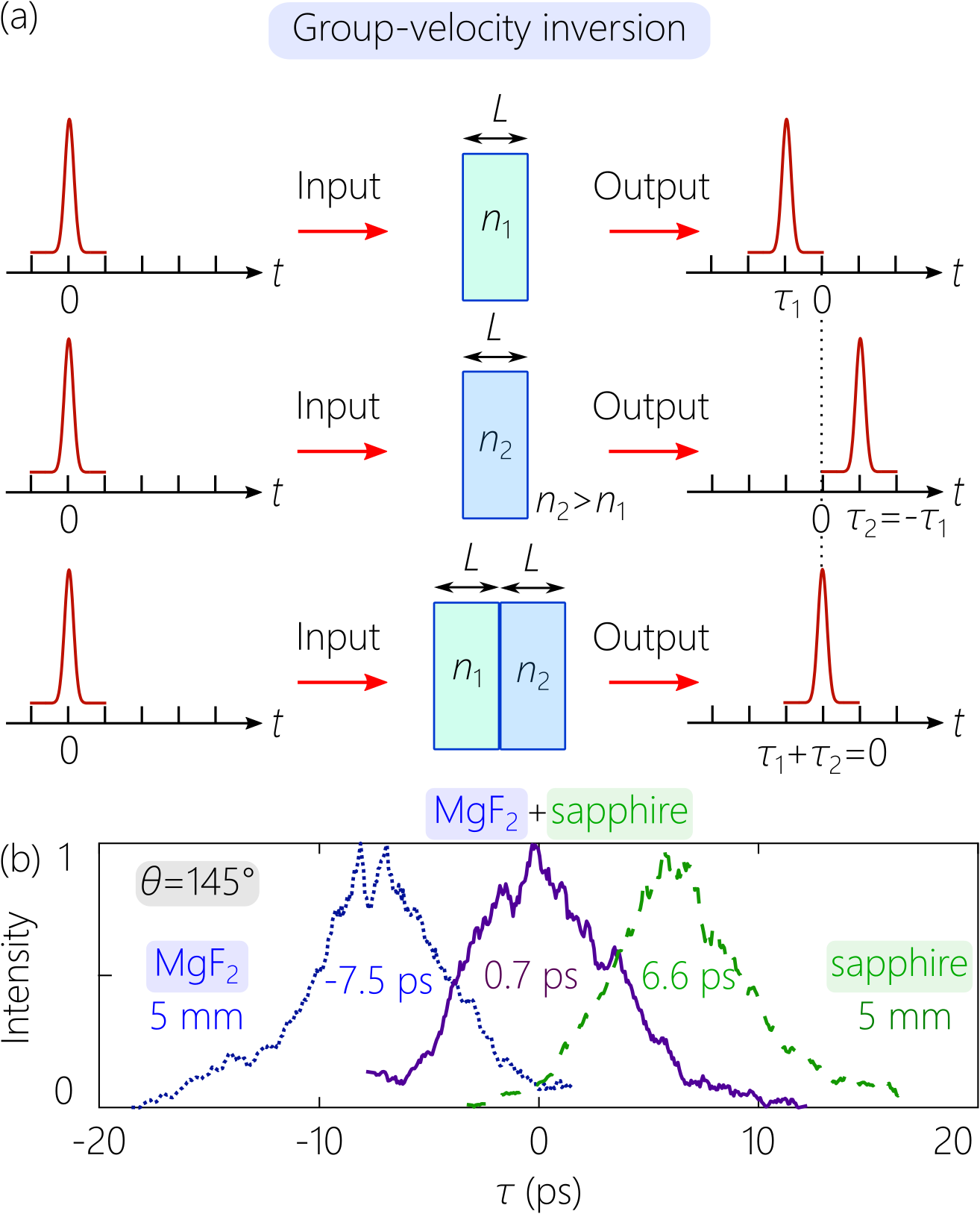}
\caption{(a) Concepts of group-velocity inversion and group-delay cancellation. (b) Measured on-axis temporal profile of the wave packets after traversing a 5-mm-thick layer of MgF$_2$ (left, dotted curve), a 5-mm-thick layer of sapphire (right, dashed curve), and after a bilayer of them (center, solid curve). The incident wave packet from free space has $\widetilde{n}_{0}\!=\!-1.42$ ($\theta_{0}\!\approx\!145^{\circ}$). We indicate the measured group delay for each wave packet.}
\label{Fig:GroupVelocityInversion}
\end{figure}

In addition to group-velocity invariance ($\widetilde{v}_{1}\!=\!\widetilde{v}_{2}$), another refractive phenomenon arises for baseband ST wave packets in which the \textit{magnitude} of the group velocity remains invariant across the interface, but whose \textit{sign} changes; i.e., $\widetilde{v}_{2}\!=\!-\widetilde{v}_{1}$ or $\widetilde{n}_{2}\!=\!-\widetilde{n}_{1}$. We refer to this phenomenon as \textit{group-velocity inversion}. Substituting this condition into Eq.~\ref{Eq:LawOfRefraction} yields the solution,
\begin{equation}
\widetilde{n}_{1}=n_{1}-n_{2}=-\widetilde{n}_{2}.
\end{equation}
Both incident and transmitted wave packets are superluminal in  this scenario. If $n_{1}\!<\!n_{2}$, then $\widetilde{n}_{1}\!<\!0$ ($\theta_{1}\!>\!90^{\circ}$), while $\widetilde{n}_{2}\!>\!0$ ($\theta_{2}\!<\!90^{\circ}$); and \textit{vice versa} when $n_{1}\!>\!n_{2}$. We therefore expect that the group delays incurred by this ST wave packet after traversing equal-thickness layers of indices $n_{1}$ and $n_{2}$ to be equal in magnitude but opposite in sign; see Fig.~\ref{Fig:GroupVelocityInversion}(a). The group delay incurred by a wave packet across a bilayer comprising two media of indices $n_{1}$ and $n_{2}$, and corresponding thicknesses $L_{1}$ and $L_{2}$, is $\tau\!=\!\{\widetilde{n}_{1}L_{1}+\widetilde{n}_{2}L_{2}\}/c$. When $\widetilde{n}_{1}\!=\!n_{1}-n_{2}\!=\!-\widetilde{n}_{2}$, we have $\tau\!=\!(n_{1}-n_{2})(L_{1}-L_{2})/c$. Therefore, when the layers have equal thicknesses $L_{1}\!=\!L_{2}\!=\!L$, $\tau$ is reduced to zero. We denote this condition \textit{group-delay cancellation}. To achieve group-delay cancellation in this bilayer, we require a wave packet synthesized in free space with a group index $\widetilde{n}_{0}\!=\!1-n_{1}n_{2}$. See paper (I) for the necessary adjustment when $L_{1}\!\neq\!L_{2}$.

We have realized group-velocity inversion experimentally using a bilayer of equal-thickness layers ($L\!=\!5$~mm) of MgF$_2$ and sapphire [Fig.~\ref{Fig:GroupVelocityInversion}(b)]. We synthesized a ST wave packet in free space with $\widetilde{n}_{0}\!\approx\!-1.42$ ($\theta_{0}\!\approx\!145^{\circ}$). In MgF$_2$, this results in a group index of $\widetilde{n}_{1}\!\approx\!-0.385$ and in sapphire $\widetilde{n}_{2}\!\approx\!0.385$. The expected group delays in the two layers are $\pm6.42$~ps. The measured group delay incurred after traversing the MgF$_2$ layer is $\approx\!-7.5$~ps and in the sapphire layer is $\approx\!6.6$~ps. The measured group delay in the bilayer is $\approx\!0.7$~ps, thereby verifying group-delay cancellation. The discrepancy between the measured and predicted values is due to the difficulty in synthesizing a wave packet with negative group index. The required SLM phase pattern features rapid spatial variation, and we are limited by the number of SLM pixels and their size. These limitations can be overcome by instead utilizing lithographically inscribed phase plates \cite{Kondakci18OE,Bhaduri19OL}.

\section{Geometry of space-time wave packets on the light-cone}

The fascinating phenomena described above concerning the refraction of ST wave packets challenge our intuitions. Perhaps the most intriguing phenomenon is the transition from normal to anomalous refraction passing through the condition for group-velocity invariance. Because the group velocity of a ST wave packet is determined by its spectral support domain on the light-cone, these refractive phenomena indicate non-trivial dynamics of the spectral support domain with a changing refractive index. We examine here these dynamics and unveil a geometric effect that plays a key role in the refraction of ST wave packets, but is absent from conventional wave packets.

\subsection{Dynamics of the representation of a pair of monochromatic plane waves upon refraction}

\begin{figure}[t!]
\centering
\includegraphics[width=8.6cm]{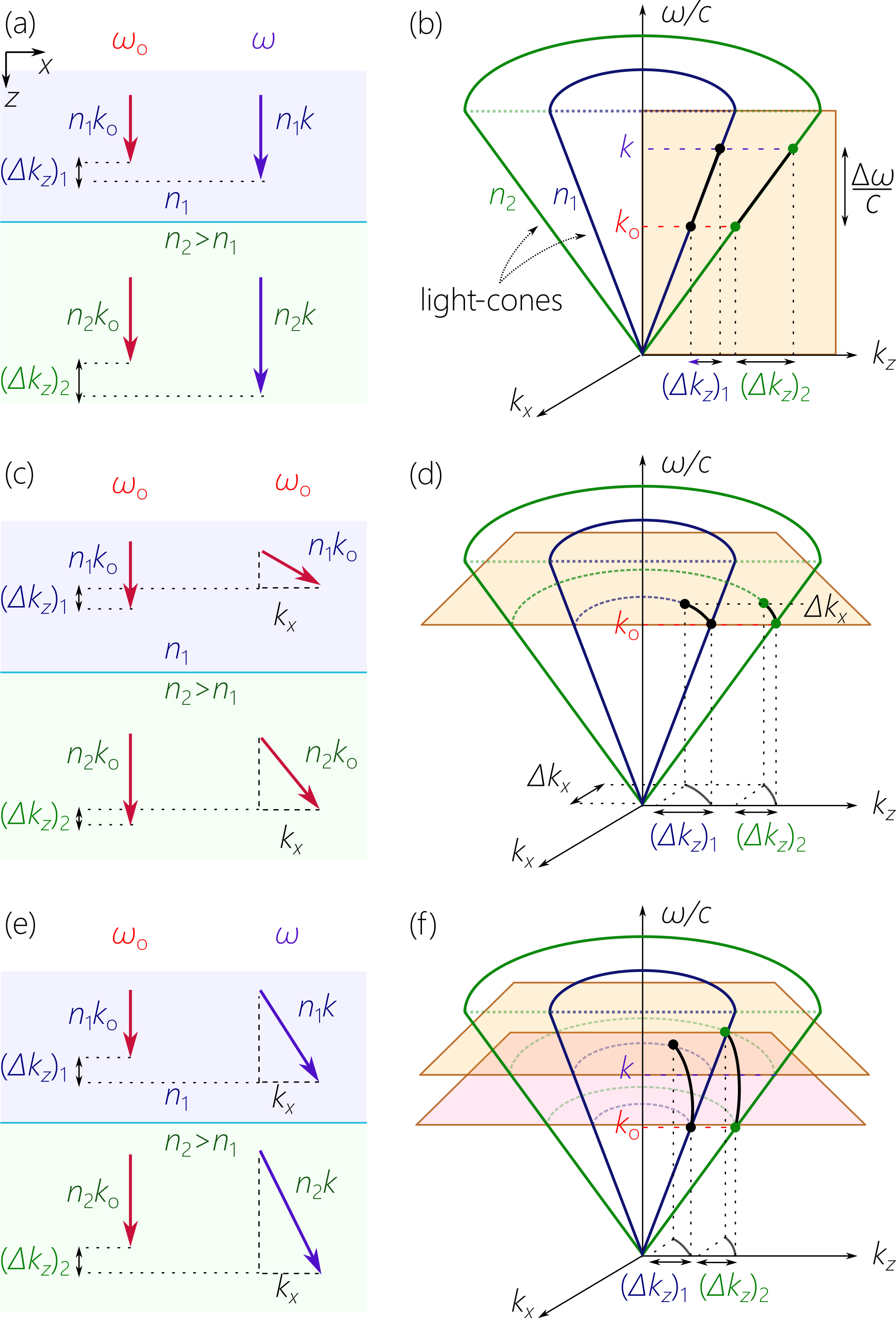}
\caption{Geometric representation of refraction of a pair of monochromatic plane waves from a medium of index $n_{1}$ to another of index $n_{2}\!>\!n_{1}$. On the left we illustrate the two wave vectors in physical space, and on the right we depict their representation as a pair of points on the light-cones associated with the two media. (a) Two plane waves at frequencies $\omega_{\mathrm{o}}$ and $\omega\!>\!\omega_{\mathrm{o}}$ are normally incident on the interface, and (b) we have $(\Delta k_{z})_{2}\!>\!(\Delta k_{z})_{1}$ because of light-cone inflation. (c) Two plane waves of the same frequency $\omega_{\mathrm{o}}$ are incident normally and obliquely on the interface, and (d) we have $(\Delta k_{z})_{2}\!<\!(\Delta k_{z})_{1}$ despite light-cone inflation. (e) Two plane waves at frequencies $\omega_{\mathrm{o}}$ and $\omega\!>\!\omega_{\mathrm{o}}$ are incident normally and obliquely, respectively, on the interface. This more general case combines (a) and (c). (f) Here we may have either $(\Delta k_{z})_{2}\!<\!(\Delta k_{z})_{1}$ or $(\Delta k_{z})_{2}\!>\!(\Delta k_{z})_{1}$.}
\label{Fig:SchematicOfThreeCases}
\end{figure}

We start by examining the dynamics of the geometric representation of a pair of monochromatic plane waves on the surface of the light-cone upon refraction [Fig.~\ref{Fig:SchematicOfThreeCases}]. We approximate the light-cone $k_{x}^{2}+k_{z}^{2}\!=\!n^{2}(\tfrac{\omega}{c})^{2}$ in a medium of index $n$ in the vicinity of $k_{x}\!=\!0$ and $\omega\!=\!\omega_{\mathrm{o}}$ for a narrowband ($\Delta\omega\!\ll\!\omega_{\mathrm{o}}$) paraxial ($\Delta k_{x}\!\ll\!k_{\mathrm{o}}$) field as:
\begin{equation}\label{Eq:ApproximateKz}
k_{z}\!\approx\!n\frac{\omega}{c}-\frac{k_{x}^{2}}{2nk_{\mathrm{o}}}.
\end{equation}
The first term is related to the temporal DoF, and the second term to the spatial DoF. Transmission from a medium of index $n_{1}$ to another of index $n_{2}\!>\!n_{1}$ is accompanied by `light-cone inflation': the surface of the light-cone is enlarged and the distance separating any pair of points on its surface increases. One thus anticipates that the separation between the projections of two such points onto the $k_{z}$-axis, $\Delta k_{z}$, will increase with light-cone inflation. The $\Delta k_{z}$ projection is crucial in determining the group velocity for a wave packet at fixed $\Delta\omega$. We show below that $\Delta k_{z}$ does \textit{not} necessarily increase with light-cone inflation for ST wave packets.

\subsubsection{Refraction of two plane waves of different temporal frequencies but the same spatial frequency}

Consider two monochromatic plane waves of different temporal frequencies $\omega_{\mathrm{o}}$ and $\omega\!>\!\omega_{\mathrm{o}}$ that are incident \textit{normally} on the interface. Both plane waves lie on the light-line $k_{z}\!=\!nk$, where $k\!=\!\omega/c$ [Fig.~\ref{Fig:SchematicOfThreeCases}(a,b)]. These two plane waves can be taken as the end points of the spectrum of a plane-wave pulse with temporal bandwidth $\Delta\omega\!=\!\omega-\omega_{\mathrm{o}}$ [Fig.~\ref{Fig:SchematicOfThreeCases}(b)]. Therefore, $\Delta k_{z}\!=\!k_{z}(\omega)-k_{z}(\omega_{\mathrm{o}})$ increases from $(\Delta k_{z})_{1}\!=\!n_{1}\tfrac{\Delta\omega}{c}$ in the first medium to $(\Delta k_{z})_{2}\!=\!n_{2}\tfrac{\Delta\omega}{c}$ in the second, so that $(\Delta k_{z})_{2}\!>\!(\Delta k_{z})_{1}$ when $n_{2}\!>\!n_{1}$. Our traditional expectation is thus met, and light-cone inflation results in an increase in $\Delta k_{z}$ and a drop in $\widetilde{v}\!=\!\tfrac{\Delta\omega}{\Delta k_{z}}$ as usual upon travel from low-index to high-index.

\subsubsection{Refraction of two plane waves of the same temporal frequency but different spatial frequencies}

Consider now two plane waves of the \textit{same} temporal frequency $\omega_{\mathrm{o}}$, one incident normally ($k_{x}\!=\!0$) and the other obliquely ($k_{x}\!\neq\!0$); see Fig.~\ref{Fig:SchematicOfThreeCases}(c,d). These two plane waves can be taken to be the end points of the spatial spectrum of a monochromatic beam (with $k_{x}$ for the second plane wave corresponding to the spatial bandwidth $\Delta k_{x}$ of the beam) [Fig.~\ref{Fig:SchematicOfThreeCases}(d)]. Here, $(\Delta k_{z})_{1}\!\approx\!\tfrac{(\Delta k_{x})^{2}}{2n_{1}k_{\mathrm{o}}}$, $(\Delta k_{z})_{2}\!\approx\!\tfrac{(\Delta k_{x})^{2}}{2n_{2}k_{\mathrm{o}}}$, and thus $(\Delta k_{z})_{2}\!<\!(\Delta k_{z})_{1}$ when $n_{2}\!>\!n_{1}$. In contrast to the preceding case of a plane-wave pulse, light-cone inflation results in \textit{shrinkage} of $\Delta k_{z}$ for a monochromatic beam. This shrinkage is a consequence of the constraint imposed by the invariance in $\Delta k_{x}$ across the planar interface [Fig.~\ref{Fig:SchematicOfThreeCases}(d)]. Although the phase velocity drops from $c/n_{1}$ to $c/n_{2}$, we cannot however define a group velocity for a monochromatic beam. The $\Delta k_{z}$ shrinkage will nevertheless be crucial in determining the change in the group velocity of ST wave packets.

\subsubsection{Refraction of two plane waves of different spatial and temporal frequencies}

We now combine the two cases examined above and consider two plane waves of temporal frequencies $\omega_{\mathrm{o}}$ and $\omega\!>\!\omega_{\mathrm{o}}$, with the first normally incident ($k_{x}\!=\!0$) and the second obliquely incident ($k_{x}\!\neq\!0$) on the interface [Fig.~\ref{Fig:SchematicOfThreeCases}(e,f)]. Here $(\Delta k_{z})_{j}\!\approx\!n_{j}\tfrac{\Delta\omega}{c}+\tfrac{(\Delta k_{x})^{2}}{2n_{j}k_{\mathrm{o}}}$, $j\!=\!1,2$, where the first term (associated with normally incident plane waves of different temporal frequencies) increases with $n$ because of light-cone inflation, and the second term (associated with iso-frequency plane waves incident at different angles) drops with increasing $n$ because of the invariance of $\Delta k_{x}$. Consequently, the dependence of $\Delta k_{z}$ on $n$ is no longer monotonic, which opens the door for the possibility of anomalous refraction. 

\subsection{The spectral curvature}

\begin{figure}[t!]
\centering
\includegraphics[width=8cm]{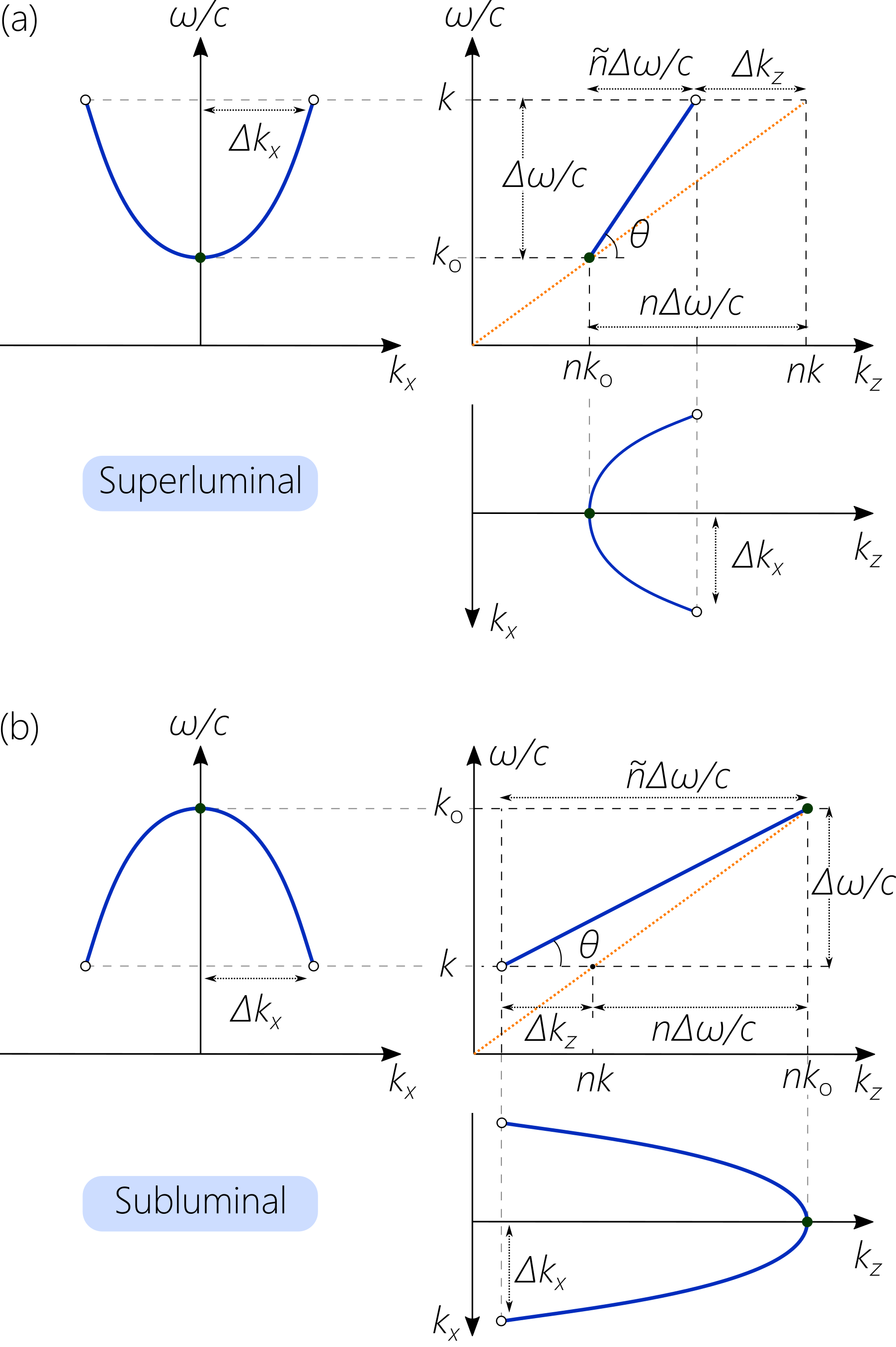}
\caption{Geometric representation of the spectral projections of an ST wave packet onto the $(k_{x},\tfrac{\omega}{c})$, $(k_{z},\tfrac{\omega}{c})$, and $(k_{x},k_{z})$ planes for (a) superluminal and (b) subluminal ST wave packets.}
\label{Fig:ProjectionsForSTWP}
\end{figure}

Each spatial frequency $k_{x}$ in a ST wave packet is uniquely related to a temporal frequency $\omega$. Taking the two plane waves in Fig.~\ref{Fig:SchematicOfThreeCases}(e,f) to be the end points of the ST wave-packet spectrum, a one-dimensional curve connects the corresponding points on the light-cone surface [Fig.~\ref{Fig:ProjectionsForSTWP}]. This curve is the conic section at the intersection of the light-cone with a tilted plane that is parallel to the $k
_{x}$-axis. The projection of this curve onto the $(k_{z},\tfrac{\omega}{c})$-plane is thus a straight line, with $\widetilde{v}\!=\!\tfrac{\Delta\omega}{\Delta k_{z}}$ and $\widetilde{n}\!=\!c\tfrac{\Delta k_{z}}{\Delta\omega}$. Because $\Delta\omega$ and $\Delta k_{x}$ are invariant, variation in $\widetilde{v}$ upon refraction are due to changes in $\Delta k_{z}$. Anomalous refraction of ST wave packets indicates that in some scenarios $\Delta k_{z}$ unexpectedly decreases rather than increases upon transmission across a planar interface to a higher-index medium.
 
Starting from Eq.~\ref{Eq:ApproximateKz}, we define $\Delta k_{z}\!=\!k_{z}(\omega)-k_{z}(\omega_{\mathrm{o}})\!=\!=\!n\frac{\Delta\omega}{c}-\tfrac{(\Delta k_{x})^{2}}{2nk_{\mathrm{o}}}$, so that the group index $\widetilde{n}$ is given by:
\begin{equation}\label{Eq:GroupIndex}
\widetilde{n}\!=\!n-\frac{g}{n}.
\end{equation}
The first term in $\widetilde{n}$ is related to light-cone inflation, while the second is a contribution from the shrinkage in $\Delta k_{z}$ due to the tight association between spatial and temporal frequencies unique to ST wave packets. The quantity $g$ is defined as:
\begin{equation}\label{Eq:EquationForG}
g=\frac{1}{2}\left(\frac{\Delta k_{x}}{k_{\mathrm{o}}}\right)^{2}\frac{\omega_{\mathrm{o}}}{\Delta\omega},
\end{equation}
which is related to the curvature of the approximate parabolic spectral projection onto the $(k_{x},\tfrac{\omega}{c})$-plane, and is independent of $n$. We therefore denote this quantity the \textit{spectral curvature}, which is positive for superluminal ST wave packets and negative for subluminal ones; see Fig.~\ref{Fig:ProjectionsForSTWP}(a,b).

\begin{figure}[t!]
\centering
\includegraphics[width=8cm]{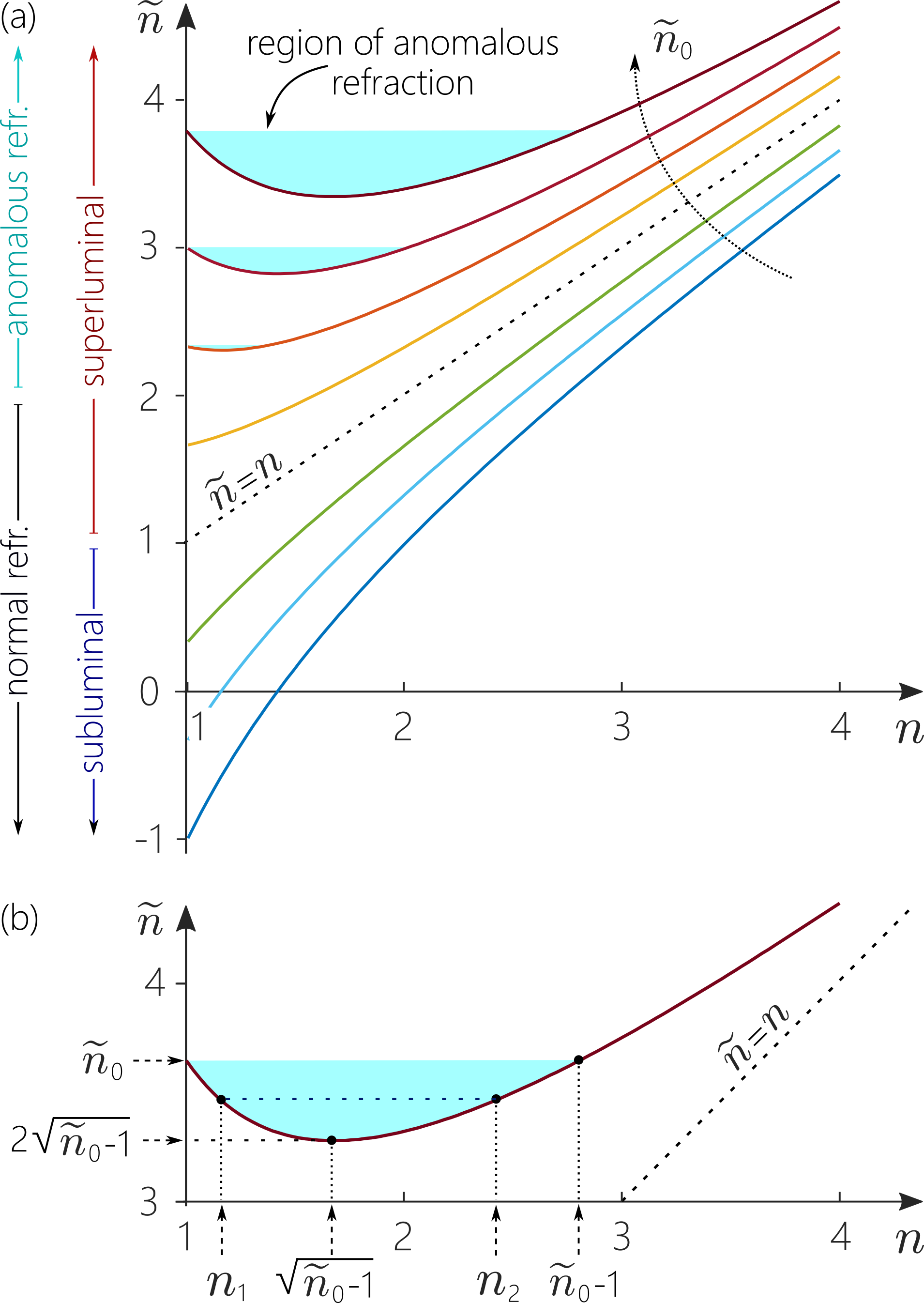}
\caption{(a) Plot of the group index $\widetilde{n}$ of a ST wave packet as a function of the refractive index $n$ for a given value of $\widetilde{n}_{0}$, according to Eq.~\ref{Eq:LawOfRefraction}. Each curve represents the class of ST wave packets that share the same value of $g$, and is identified by $\widetilde{n}_{0}\!=\!1-g$. The dotted line corresponds to a luminal plane-wave pulse $\widetilde{n}\!=\!n$ ($\widetilde{n}_{0}\!=\!1$). (b) Sketch of the evolution of $\widetilde{n}$ with $n$ for a ST wave packet of fixed $\widetilde{n}_{0}\!=\!3.6$.}
\label{Fig:GroupIndexPlot}
\end{figure}

Because $g$ is a quantity that characterizes the ST wave packet but is independent of $n$, it identifies not one ST wave packet, but rather an equivalency class of ST wave packets. If we choose as a reference the ST wave packet in free space $n_{0}\!=\!1$, whereupon $\widetilde{n}_{0}\!=\!1-g$, then the group index for a ST wave packet that shares the same value of $g$ but propagates in a medium of index $n$ is given by [Fig~\ref{Fig:GroupIndexPlot}(a)]:
\begin{equation}\label{Eq:nTilde}
\widetilde{n}=n+\frac{\widetilde{n}_{0}-1}{n}.
\end{equation}
Therefore $g$ identifies the class of all ST wave packets whose group indices satisfy this relationship in any medium of index $n$. Furthermore, a ST wave packet that is \textit{superluminal} in free space $\widetilde{n}_{0}\!<\!1$ remains so in any other medium $\widetilde{n}\!<\!n$, and similarly for \textit{subluminal} wave packets. Moreover, $\widetilde{n}\!\rightarrow\!n$ at large $n\!\gg\!\widetilde{n}_{0}$, whereupon the ST wave packet approaches a conventional one.

\begin{figure*}[t!]
\centering
\includegraphics[width=17.5cm]{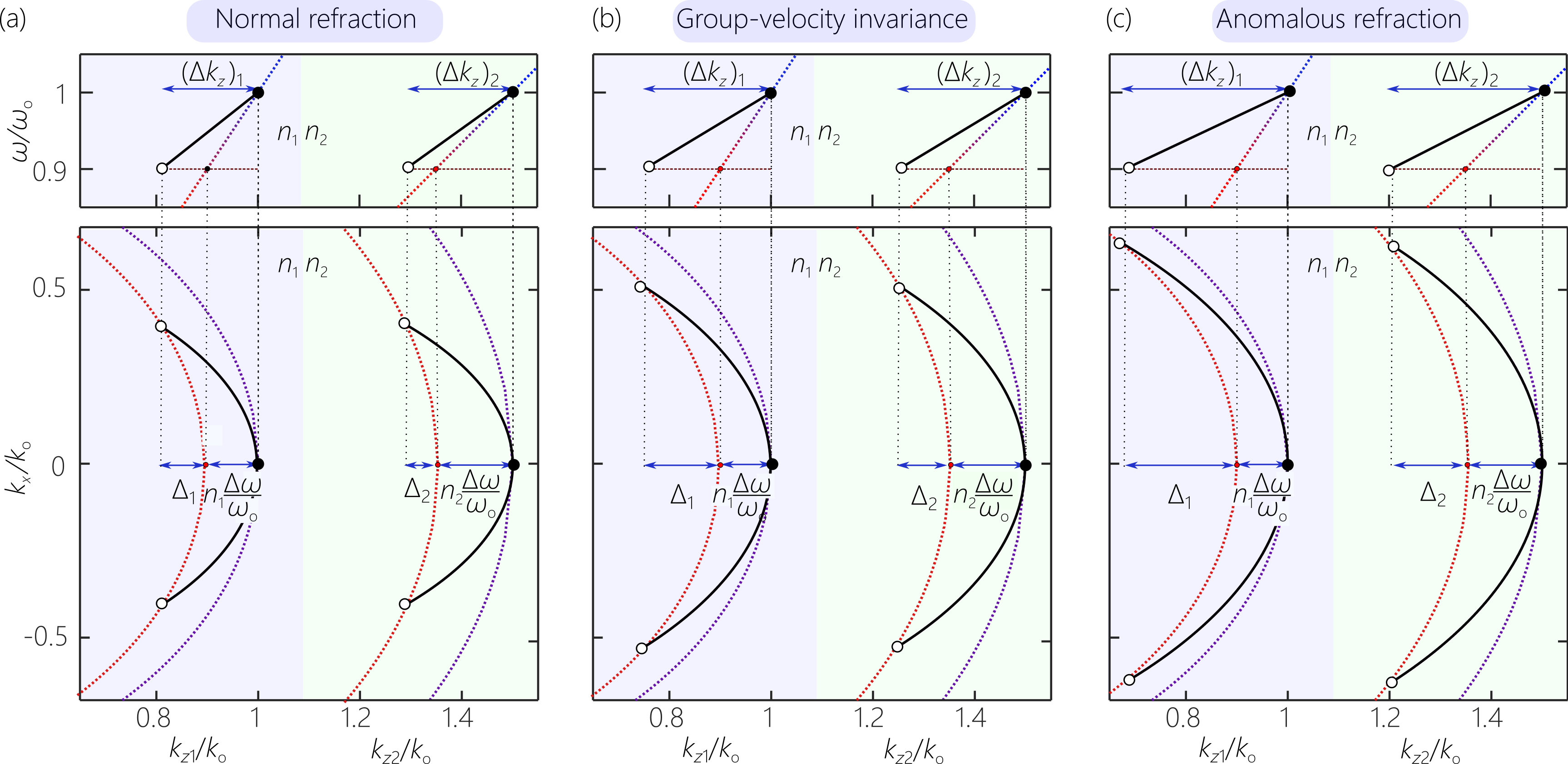}
\caption{Geometric representation of a ST wave packet in the (a) normal refraction regime, (b) group-velocity invariance regime and (c) anomalous refraction regime.  Upper row corresponds to the projection onto the $(k_{z},\omega/c)$-plane and lower to the projection onto the $(k_{x},k_{z})$-plane.}
\label{Fig:OriginOfAnomalousProjection}
\end{figure*}

\subsection{Anomalous refraction}

Normal refraction refers to the usual increase in $\widetilde{n}$ with increasing $n$, whereas anomalous refraction refers to the opposite trend of $\widetilde{n}$ \textit{decreasing} with increasing $n$. From a physical standpoint, anomalous refraction corresponds to those cases where shrinkage in $\Delta k_{z}$ resulting from the invariance in $\Delta k_{x}$ dominates over the expansion in $\Delta k_{z}$ stemming from light-cone inflation. Group-velocity invariance occurs when these two effects exactly balance each other. It is clear from Eq.~\ref{Eq:GroupIndex} that $\widetilde{n}$ decreases with increasing $n$ only when $g\!<\!-1$; i.e., anomalous refraction is only exhibited by subluminal ST wave packets for which $\widetilde{n}_{0}\!>\!2$, whereas subluminal wave packets for which $1\!<\!\widetilde{n}_{0}\!<\!2$ and all superluminal wave packets experience only normal refraction.

Let us follow the variation in $\widetilde{n}$ for a subluminal ST wave packet belonging to an equivalency class with $\widetilde{n}_{0}\!>\!2$ as we gradually increase the refractive index $n$. For concreteness, we consider ST wave packets sharing the same value of $g$ corresponding to $\widetilde{n}_{0}\!=\!3.6$; see Fig.~\ref{Fig:GroupIndexPlot}(b). Initially, when $n_{0}\!=\!1$, we have $\widetilde{n}\!=\!\widetilde{n}_{0}$. As $n$ increases, $\widetilde{n}$ first decreases, thereby exhibiting anomalous refraction. The group index $\widetilde{n}$ continues to decrease with $n$ until it reaches a minimum value of $\widetilde{n}_{\mathrm{min}}\!=\!2\sqrt{\widetilde{n}_{0}-1}$ when $n\!=\!\sqrt{\widetilde{n}_{0}-1}$, after which $\widetilde{n}$ increases with $n$ until it regains its initial value $\widetilde{n}\!=\!\widetilde{n}_{0}$: i.e., a medium of index $n\!>\!n_{0}\!=\!1$ in which nevertheless the group index of the wave packet is the same as that in free space \cite{Bhaduri19Optica}. This degeneracy ($\widetilde{n}\!=\!\widetilde{n}_{0}$ when $n\!\neq\!n_{0}$) occurs for a medium whose index is a solution to the quadratic equation $n^{2}-n\widetilde{n}+(\widetilde{n}_{0}-1)\!=\!0$ after setting $\widetilde{n}\!=\!\widetilde{n}_{0}$, from which we obtain $n\!=\!\widetilde{n}_{0}-1$. Upon normal incidence from free space onto this medium, the wave packet would retain the value of its group velocity. Subsequently, $\widetilde{n}$ increases monotonically with $n$.

Consider now subluminal ST wave packets belonging to the same equivalency class (constant $g$) normally incident from a medium of index $n_{1}$ to another of index $n_{2}\!>\!n_{1}$, so that $\widetilde{n}_{1}\!=\!n_{1}-g/n_{1}$ and $\widetilde{n}_{2}\!=\!n_{2}-g/n_{2}$. The condition for group-velocity invariance requires that $\widetilde{n}_{1}\!=\!\widetilde{n}_{2}$ in these two media, which identifies a unique equivalency class with $\widetilde{n}_{0}\!=\!1+n_{1}n_{2}\!>\!2$, whereby $\widetilde{n}_{1}\!=\!\widetilde{n}_{2}\!=\!n_{1}+n_{2}$. As we increase $n$ above $n_{1}$, the group index $\widetilde{n}$ decreases below $\widetilde{n}_{1}\!=\!n_{1}+n_{2}$, thus exhibiting anomalous refraction. The group index reaches a minimum value of $\widetilde{n}_{\mathrm{min}}\!=\!2\sqrt{n_{1}n_{2}}$ at and $n\!=\!\sqrt{n_{1}n_{2}}$. With further increase in $n$, $\widetilde{n}$ increases and regains its initial value of $\widetilde{n}_{2}\!=\!\widetilde{n}_{1}$ when $n\!=\!n_{2}$ and subsequently increases monotonically.

\subsection{ST wave packets on a light-cone with increasing group index}

An alternative perspective on the dynamics of the spectral support domain for a normally incident ST wave packet can be gleaned by considering two media of \textit{fixed} indices $n_{1}$ and $n_{2}\!>\!n_{1}$ as we tune $\widetilde{v}_{1}$ while holding $\Delta\omega$ fixed. In other words, we hold the refractive indices fixed and change the equivalency class (tune $g$) by varying the spatial bandwidth $\Delta k_{x}$. We plot in Fig.~\ref{Fig:OriginOfAnomalousProjection} the spectral projection onto the $(k_{z},\tfrac{\omega}{c})$-plane to highlight the $\omega$-invariance, and onto the $(k_{x},k_{z})$-plane to highlight the $k_{x}$-invariance.

Consider a ST wave packet having temporal and spatial bandwidths of $\Delta\omega$ and $\Delta k_{x}$, respectively. Because $\Delta\omega$ and $\Delta k_{x}$ are invariant with refraction, we have:
\begin{equation}\label{Eq:DifferenceInAxialBandwidths}
(\Delta k_{z})_{2}-(\Delta k_{z})_{1}\!=\!(n_{2}-n_{1})\frac{\Delta\omega}{c}+\frac{(\Delta k_{x})^{2}}{2k_{\mathrm{o}}}\left(\frac{1}{n_{2}}-\frac{1}{n_{1}}\right);
\end{equation}
$(\Delta k_{z})_{2}-(\Delta k_{z})_{1}\!=\!0$ corresponds to group-velocity invariance; $(\Delta k_{z})_{2}-(\Delta k_{z})_{1}\!>\!0$ to normal refraction; and $(\Delta k_{z})_{2}-(\Delta k_{z})_{1}\!<\!0$ to anomalous refraction. In conventional optical fields whose spatial and temporal DoFs are separable, $\Delta\omega$ and $\Delta k_{x}$ are independent of each other. In contrast, $\Delta\omega$ and $\Delta k_{x}$ are related in ST wave packets through the spectral curvature $g$, so that:
\begin{equation}
(\Delta k_{z})_{2}-(\Delta k_{z})_{1}\!=\!(n_{2}-n_{1})\frac{\Delta\omega}{c}\left(1+\frac{g}{n_{1}n_{2}}\right).
\end{equation}
The sign of $(\Delta k_{z})_{2}-(\Delta k_{z})_{1}$ depends on the value of $g$.

We start in Fig.~\ref{Fig:OriginOfAnomalousProjection}(a) with a subluminal ST wave packet in the normal-refraction regime, and $\widetilde{n}_{1}\!<\!n_{1}+n_{2}$ or $g\!>\!-n_{1}n_{2}$. As the wave packet traverses the interface to the higher-index medium, light-cone inflation enhances the first term in Eq.~\ref{Eq:DifferenceInAxialBandwidths} beyond the drop in the second, so that $(\Delta k_{z})_{2}\!>\!(\Delta k_{z})_{1}$ and $\widetilde{v}_{2}\!<\!\widetilde{v}_{1}$, thereby producing normal refraction. We now increase $\widetilde{n}_{1}$ at fixed $\Delta\omega$, which necessitates increasing $\Delta k_{x}$ [Fig.~\ref{Fig:OriginOfAnomalousProjection}(b)]. The same change in index from $n_{1}$ to $n_{2}$ leads to the same increase in the first term of Eq.~\ref{Eq:DifferenceInAxialBandwidths} as in the first case [Fig.~\ref{Fig:OriginOfAnomalousProjection}(a)]. However, the increase in $\Delta k_{x}$ entails that the magnitude of the drop in the second term in Eq.~\ref{Eq:DifferenceInAxialBandwidths} exceeds that in the previous case. At $g\!=\!-n_{1}n_{2}$, the increase in the first term in Eq.~\ref{Eq:DifferenceInAxialBandwidths} balances the decrease in the second term. Consequently, $(\Delta k_{z})_{1}\!=\!(\Delta k_{z})_{2}$, and group-velocity invariance is reached. Further increasing $\widetilde{n}_{1}$ by increasing $\Delta k_{x}$ at fixed $\Delta\omega$ so that $g\!<\!-n_{1}n_{2}$ results in an overall decrease in $\Delta k_{z}$, $(\Delta k_{z})_{2}\!<\!(\Delta k_{z})_{1}$ [Fig.~\ref{Fig:OriginOfAnomalousProjection}(c)]. Consequently $\widetilde{v}_{2}\!>\!\widetilde{v}_{1}$, and we enter the anomalous refraction regime.

\section{Discussion and Conclusion}

The fascinating refractive phenomena reported here for ST wave packets are all a consequence of their precise spatio-temporal structure, which is a manifestation of `classical entanglement' \cite{Qian11OL,Kagalwala13NP}. In previous studies, classical entanglement has been exclusively studied with regards to pairs of \textit{discrete} optical degrees of freedom; e.g., polarization, two spatial modes, or two temporal modes (or time bins) \cite{Qian11OL,Kagalwala13NP,Abouraddy14OL,Kagalwala15SR,Berg15Optica,Aiello15NJP}. Here classical entanglement is realized between two \textit{continuous} degrees of freedom: the spatial frequencies $k_{x}$ and the temporal frequencies $\omega$. This correlation undergirds the unique features of ST wave packets, including propagation invariance \cite{Kondakci17NP,Schepler20ACSP}, self-healing \cite{Kondakci18OL}, and time-diffraction \cite{Kondakci18PRL,Yessenov20PRL}. Furthermore, a `degree of classical entanglement' can be defined that determines the wave packet propagation distance when the association between each $k_{x}$ and $\omega$ is not ideally strict \cite{Kondakci19OL}. Our work here reveals that classical entanglement also impacts the refraction of ST wave packets. Indeed, the dynamics of this classical entanglement between the spatial and temporal DoFs on the surface of the light-cone undergird the refractive phenomena we have reported here.

The key to the refraction of ST wave packets is the tunability of their group velocity independently of the refractive index of the medium \cite{Bhaduri19Optica,Bhaduri20NP}. This warrants drawing a comparison to the work on `slow-light' and `fast-light' \cite{Boyd09Science} where the group velocity is modified via strong linear chromatic dispersion produced typically by a spectrally narrow resonance in an optical material or a photonic structure. Crucially, the accessible bandwidth for such pulses are usually extremely narrow because of this resonant nature. Moreover, such resonances introduce either strong optical losses or amplification. To date, slow- and fast-light have been realized without manipulation of the spatial DoF, so that refraction is unlikely to produce significant changes. In contrast, our non-resonant approach relies on spatio-temporal coupling to introduce angular dispersion that is responsible for tailoring the group velocity. Consequently, the propagation and refraction of ST wave packets can be realized and observed in traditional passive optical materials (or planar waveguides \cite{Shiri20NC}, and there are no limits -- in principle -- on the exploitable bandwidth \cite{Yessenov20NC}. 

Finally, recent work has examined `temporal refraction' where changes in the refractive index of a medium occur in time while retaining spatial homogeneity \cite{Plansinis15PRL}, which typically requires a nonlinear effect for its realization. In contrast, the spatial changes associated with traditional refraction produce the effects verified here with ST wave packets in linear media.

In conclusion, we have provided experimental demonstrations of the remarkable refractive effects associated with basedband ST wave packets at normal incidence on a planar interface between a pair of non-dispersive, homogeneous, isotropic dielectrics. We have confirmed the predicted phenomena of group-velocity invariance, anomalous refraction, and group-velocity inversion (and its corollary, group-delay cancellation). These phenomena do not occur with any other optical field configuration in non-dispersive media. We have realized these effects in generic optical materials with refractive indices in the range from 1.38 to 1.76, with no fundamental limit on the wave packet bandwidth. These findings reinforce the notion that ST wave packets should be considered as objects in their own right identified by an internal degree of freedom (the spectral tilt angle or spectral curvature). These results are critical for furthering our understanding of the interaction of ST wave packets with photonic devices.

\section*{Funding}
U.S. Office of Naval Research (ONR) contract N00014-17-1-2458.

\vspace{2mm}
\noindent
\textbf{Disclosures.} The authors declare no conflicts of interest.

\bibliography{diffraction}

\end{document}